\documentclass[12pt,aps,preprint,epsfig,epsf]{aastex}

\slugcomment{{}}

\begin{document}
\title{Relaxation-limited evaporation of globular clusters}
\author{Maurice H.P.M. van Putten}
\affil{Korea Institute for Advanced Study, Dongdaemun-Gu, Seoul 130-722, Korea}

\begin{abstract}
Evaporative evolution of stellar clusters is shown to be relaxation limited when the number of stars satisfies $N>>N_c$, where $N_c\simeq 1600$. {For a Maxwell velocity distribution that extends beyond the escape velocity,} this process is {\em bright} in that the Kelvin-Helmholtz time scale, $f_H^{-1}t_{relax}$, is shorter than the Ambartsumian-Spitzer time scale, $f_N^{-1}t_{relax}$, where $f_H>f_N$ denote the fractional changes in total energy and number of stars per relaxation time, $t_{relax}$. The resulting evaporative lifetime $t_{ev}\simeq 20.5 t_{relax}$ for isolated clusters is consistent with Fokker-Planck and N-body simulations, where $t_{relax}$ is expressed in terms of the half-mass radius. We calculate the grey body factor by averaging over the anisotropic perturbation of the potential barrier across the tidal sphere, and derive the tidal sensitivity ${d\ln t_{ev}}/{dy}\simeq -1.9$ to $-0.7$ as a function of the ratio $y$ of the virial-to-tidal radius. Relaxation limited evaporation applies to the majority of globular clusters of the Milky Way with $N=10^4-10^6$ that are in a pre-collapse phase. It drives streams of stars into the tidal field with a mean kinetic energy of 0.71 relative to temperature of the cluster. Their $S$ shape morphology leads in sub-orbital and a trails in super-orbital streams separated by $3.4\sigma/\Omega$ in the radial direction of the orbit, where $\Omega$ denotes the orbital angular velocity and $\sigma$ the stellar velocity dispersion in the cluster. These correlations may be tested by advanced wide field photometry and spectroscopy.
 \end{abstract}
 
\section{Introduction}

Globular clusters represent some of the oldest stellar systems as indicated by their stellar ages of at least 12 Gyr \citep{pee68,van96,han02} consistent with the ages of white dwarfs in the thin disk of the Milky Way \citep{noh92,sal00,del08,ise08,ren10}, the oldest halo field stars such as HE 2523-0901 of 13.2 Gyr$\pm2.7$ Gyr (13.4$\pm$1.8 Gyr based on U/r  chronometry alone, \cite{fre07}) and the age of the Universe of 13.75 $\pm 0.11$ Gyr \citep{jar10}. The population of clusters of the Milky Way as observed today is but a shadow of a primordial population \citep{che90,mcl08,pri08}, that may have formed in the era of reionization \citep{van96}. Their ages hereby exceed their relaxation times by a factor of a few to an order of magnitude. They are essentially virialized close to thermal equilibrium \citep{mey97} and some 80\% satisfy the King model (KM, \cite{kin66,che90}), which bears out well also in numerical simulations \citep{cha10}. The exception is a minor fraction of clusters in the direction of the galactic halo with a centrally peaked core, commonly designated as post core-collapse (PCC) \citep{che89,tra93,joh08}. 

Prior to core collapse, the evolution of stellar clusters is well-described by two-body interactions in the small angle scattering approximation \citep{spu96,aar98}, that naturally gives rise to evaporation of the cluster \citep{amb38,spi40}. This process is amenable to detailed modeling by Fokker-Planck and N-body simulations \citep{hen61,spu96}, there results of which can be interpreted statistically (reviewed in \cite{mey97}) in terms of evaporation, tidal effects, stellar mass loss, mass segregation and heating and crossings of the galactic disk \citep{gne97,wat00,jos01,fre02,heg01,cha10,heg11}. 

There is an increasing interest in these various processes to develop efficient semi-analytic approaches to the birth and evolution of large numbers of stellar clusters as tracers of cosmological evolution of galaxies \citep{kru11}. Evaporation stands out as a common process in many-body systems that includes for instance atmospheric science (e.g. \cite{cha63,hay65,cha67,shi86,joh08,joh10}), where it extracts the lightest elements from an otherwise nearly thermal state (e.g. \cite{cha96,cha02}). 

Evaporation is a radiation process that defines a finite Kelvin-Helmholtz time scale due to energy extraction by the escaping stars. Key to their evolution, therefore, is an accurate determination of the energy loss carried off by stars escaping the cluster a determination of the time scale of evaporation, in particular whether it follows the relaxation time scale or the crossing time scale \citep{heg01} or possibly a combination thereof \citep{bau01}. The energy extraction by escaping stars is potentially useful in classifying post core-collapse clusters \citep{kup08b}.

Evaporation of stellar clusters has received considerable attention (reviewed by \cite{joh93}) with an emphasis on mass loss typically more so than energy loss (but see \cite{wie67}). Yet, the {  energy} distribution function of all stars, those remaining in the globular cluster and those expelled by evaporation, continuously spreads towards increasingly negative and positive values associated with contraction and heating. The kinetic energies of evaporated stars in the halo hereby carry a thermal imprint of the history of the clusters. Since the {  energy} distribution function is unbounded in the point particle limit, there is the potential for extended lifetimes already at small binary mass fractions. However, observational evidence for a dynamically significant population of compact binaries has not been demonstrated, nor in the correlation between the observed population of luminous X-ray binaries and the binary star-star interaction rate \citep{poo03} that points to relatively small binary populations that do not necessarily affect the lifetime of the clusters. 

In this paper, we explore the parameter range for relaxation-limited evaporation{, the effect of energy loss,} and some observational consequences for the present state of globular clusters in the Milky Way.  We discuss the thermodynamics of evaporating globular clusters with tidal perturbation in the large-N limit of equal mass stars as a function of a tidal parameter
\begin{eqnarray}
y=\frac{R}{r_t},
\label{EQN_t03}
\end{eqnarray}
representing the ratio between the virial radius $R$ (close to the half-mass radius, to be discussed further in \S2 below) and the tidal radius $r_t$ of the cluster. For old clusters such as of the Milky Way, $r_t$ is much larger than the radius of the cluster. The tidal sphere has the effect of lowering the potential barrier for stars to escape along orbits into the the halo of the galaxy and, generally, introduces strong anisotropy in the directions of the escaping stars defined by the Lagrange points $L_1$ and $L_2$ \citep{hay67}. 
Existing approaches to incorporate tidal effects commonly use the minimum energy for escape at $L_1$ \citep{spi87,lee87,che90,tak97,ros97}, effectively applying a spherically symmetric extension thereof to the tidal sphere. However, since last scatterings arise typically from within the half-mass radius \citep{bau02}, the escape rate is effectively determined by the probability of escape derived from spherical averaging over the anisotropic tidal field for randomly oriented ejections. We thus derive the evaporative lifetime up to the moment of core-collapse to be
\begin{eqnarray}
t_{ev} \simeq 20.53  \left(1-1.90y+2.78y^2-2.40y^3+0.84y^4+O(y^5)\right)t_{relax},
\label{EQN_t04}
\end{eqnarray}
where the relaxation time $t_{relax}$ expressed in terms of the half-mass radius $R_h$ (the coefficient is $13.85$ when $t_{relax}$ is expressed in $R$). While our approach is {a largely explorative extension to \cite{amb38} focused on the sensitivity of the evaporative lifetime to energy loss, some} confidence can be gained by comparing (\ref{EQN_t04}) with known results reviewed in \cite{spi87,tak97,mey97,heg03}.

Early in their evolution over the first few Gyr, globular clusters may have experienced considerable mass loss in stellar winds \citep{aar98,bau03} due to stellar evolution (e.g. \cite{app86}), that created a crucial factor in the likelihood of survival of a cluster after its formation \citep{che90}. A common approximation used in Fokker-Planck and N-body simulations is an exponential evolution on a time scale $t_m$ \citep{aar85,che90,mey97}. The same approach is readily incorporated in a thermodynamic model as a regular perturbation provided that $t_m>>t_{relax}$. Evaporation has also been found to narrow the mass spectrum due to a relatively enhanced rate of escape of the lightest stars,  by which older clusters tend evolves similarly to a system of equal mass stars \citep{che90}. The aforementioned analogy with atmospheric science is useful here, noting that evaporation leads to an early escape of light molecules such as hydrogen and helium from the terrestrial atmosphere leaving only the relatively heavy molecules of nitrogen and oxygen to survive. Early evolution of star clusters therefore deviates from the evaporation driven evolution of single mass clusters. However, for the old clusters of the Milky Way at the present epoch, these considerations are less critical, and will not be considered here further.

As an application, we focus on high velocity stars in and around globular clusters that may provide ``real-time" snapshot of last scattering events \citep{hen61}. These last scattering events from within the half-mass radius (in the limit of isolated clusters, \cite{bau02}) and most frequently derive from the core, where the density is highest and the local relaxation time is the shortest \citep{aar74,joh93}. This includes tidal tails produced by ejections that, indirectly, serves as a diagnostic for the cluster temperature as, e.g., in NGC 6254 \citep{leo00} and 47 Tuc \citep{lan10a,lan10b}, where the velocity dispersion rises to about 60\% at the tidal radius relative to the velocity dispersion inside the cluster. Observational data on tidal tails hereby provides a signature that may be used to calibrate numerical models when applied to individual clusters, e.g., M15 \citep{gri95,dru98}, Palomar 5 \citep{ode01,mey01,deh04}, NGC 5466 \citep{bel06} and Palomar 14 \citep{sol11} and their tails by $N-$body simulations (e.g. \cite{heg01,lee06}). This may include further processes such as additional heating by past disk crossings for clusters near the center of the galaxy, but possibly less so for the outer clusters \citep{gne97,leo00}. Some of these tidal tails display an $S-$shaped morphology \citep{heg01,bel06,kro08} along the Lagrange points $L_1$ and $L_2$ on short scales, e.g., in NGC 5904 \citep{leo00,jor10},  and along the orbit of the cluster on long scales, e.g., NGC 5644 \citep{fel07}, where the leading (trailing) tail is inside (outside) the cluster orbit around the galactic center. The temperature of the clusters varies broadly, from about 1 km/s in Pal 12 \citep{leo00} to about 15 km/s in $\omega$-Centauri \citep{van06}, with evidently different prospects for the manifestation of these tidal tails. 

The planned Large Synoptic Survey Telescope \citep{lsst09}, the Baryon Oscillation Spectroscopic Survey (BigBOSS, a multi-object spectrograph, \cite{sch09}) the Japan Astrometry Satellite Mission for INfrared Exploration (JASMINE \cite{gou10}), and GAIA \citep{lin95} seem to be ideally suited for studying these tidal tails by large field of view photometry and spectroscopy. This development offers novel opportunities for testing the physics of evaporation by measuring correlations between the kinematics of stars in tails to the thermal state of the clusters. 

In \S2, we define notation and the basic quantities describing the thermodynamic state of virialized star clusters, their crossing and relaxation time. In \S3, we introduce the general set-up for the evolution equations. in \S4, the parameter range for relaxation-limited evaporation is identified. In \S5, we calculate the evaporation coefficients for the number of stars and energy emitted per unit of relaxation time. In \S6, we extend our model by regular perturbation analysis to include tidal interactions. In \S7, we discuss the remaining relaxation-limited evaporative lifetime of the globular clusters in the Milky Way.  We give model predictions for observational test on tidal tails in \S8 and summarize our results in \S9.

\section{Virialized star clusters}

By the large number $N=10^4-6$ of stars, the globular clusters of the Milky Way are essentially virialized with constant mass-to-light ratio in their cores \citep{mey97}. For globular clusters of the Milky Way, corrections to the Virial Theorem due to tidal forces are generally small \citep{lee69}. In light of the aforementioned KM morphology, we begin with a homogeneous and isolated cluster of single stars of the same mass, $m$, and no binaries.  The system is then described by a total number of stars, $N$ and virial radius, $R$. The Newtonian binding energies $U_{ij}=-Gm^2/|r_i-r_j|$ between stars $i$ and $j$ ($i\ne j$) give rise to a total energy $H$ representing the sum of total potential energy $U$ between the stars and the total kinetic energy $E_k$ of their individual motions, satisfying 
\begin{eqnarray}
H= U+E_k=\frac{1}{4} \Sigma_{i\ne j} U_{ij} \simeq -\frac{GN^2m^2}{4R},
\label{EQN_H}
\end{eqnarray}
where we used the Virial Theorem, $2\bar{E}_k+\bar{U}=0$, on the mean values
\begin{eqnarray}
\bar{U} = \frac{1}{2N} \Sigma_{i\ne j} U_{ij} \simeq -\frac{GmM}{2R},~~\bar{E}_k=N^{-1}E_k,
\label{EQN_U}
\end{eqnarray}
{and where (\ref{EQN_H}) is a defining relationship for the virial radius $R$} (cf. \cite{mey97}). The ratio $M/R$ is a hereby a measure for the mean kinetic energy in the system.

The relaxation time for a homogeneous population of stars arises out of elastic 2-body encounters, produce random deflection angles $\Delta \phi \simeq \pm U_{ij}/E_k$, where $U_{ij}=-Gm^2/b$ denotes the binding energy associated with the impact parameter $b$ for an interaction between stars $i$ and $j$. High velocity encounters hereby experience relatively small angular deflections, giving rise to asymptotic freedom generic to scattering problems.  In a crossing time 
\begin{eqnarray}
t_c=\frac{R}{v}=\frac{R^{3/2}}{\sqrt{GmN}}, 
\label{EQN_tc}
\end{eqnarray}
up to a factor of unity, the numerous encounters of a star wandering through a cluster give a non-zero expectation value of the variance of these deflection angles. (The crossing time corresponds to the free-fall time scale within a factor of two.) The result can be summarized in a relaxation time for an expectation of unit variance \citep{spi40}
\begin{eqnarray}
t_{relax} = \frac{N}{6\ln N} t_c = 2.4  \times \frac{N^{1/2}R_1^{3/2}}{\ln N}~\mbox{Myr},
\label{EQN_tr}
\end{eqnarray}
where $R=R_1$ pc. The approximations involved allow for slight variations, while preserving the main dependencies on the overall characteristics of the stellar system. The details of these differences shall not concern us here. 

The virial radius $R$ is closely related to the half-mass radius $r_h$ are closely related, e.g., based on the Plummer model, we have \citep{aar80,mey97}
\begin{eqnarray}
\frac{GM_c^2}{2R}= -\bar{U} =\frac{3\pi GM^2_c}{32 N(r_h/1.30)}=0.38\frac{ GM_c^2}{r_h},
\label{EQN_Rr}
 \end{eqnarray} 
whereby $R=1.30 r_h$. Consequently, $t_{rh}\simeq 0.67 t_{relax}$, where $t_{rh}$ denotes the relaxation time estimated on the basis of $r_h$. 

\section{Equations for evaporative loss of mass and energy}

Evaporation is a radiation process in which stars with sufficient {  energy} escape, out to infinity for isolated clusters or out into the tidal field of the host galaxy. This process can be described by fractional changes $f_N$ and $f_H$ in the number of stars, $N$, and, respectively, the energy of the cluster per unit of a characteristic time, that is generally close to the relaxation time. It represents the statistical outcome of a relatively large phase space of possible stellar trajectories, that emerges from, most frequently, soft two-body interactions. 

When the relaxation time is intermediate between a dynamical crossing time and the evaporative lifetime of the cluster, the escape probabilities can be calculated in a thermodynamic limit subject to the Virial Theorem and radiation boundary conditions. In the absence of thermodynamic equilibrium, the ratio $f_H/f_N$ is anywhere from zero, e.g., when all stars are trapped or escape with at most the escape velocity, to arbitrarily large, e.g., in the unlikely event of a star escaping with a kinetic energy on the order of the total gravitational binding energy of the cluster. 

\cite{amb38} described evaporation of isolated clusters by a singular perturbation away from thermal equilibrium given by the tail of a Boltzmann velocity distribution representing a flux of stars with positive {energy} to escape to infinity on the relaxation time scale $t_{relax}$, that is,
\begin{eqnarray}
\frac{dN}{dt}=-f_N N t_{relax}^{-1},
\label{EQN_fN0}
\end{eqnarray}
where $f_N\simeq1/135$. (Here, $f_N$ is sometimes referred to as $\xi_e$, e.g., \cite{spi87}). Thus, (\ref{EQN_fN0}) defines the Ambartsumian-Spitzer time scale 
\begin{eqnarray}
 t_{AS} =  f_N^{-1} t_{relax}.
 \label{EQN_fev}
 \end{eqnarray}
The true evaporative lifetime of cluster is shorter than (\ref{EQN_fev}), however, by continuous heating of the cluster in a fully nonlinear evolution, upon including the energy radiated away by the escaping stars. In addition to $t_{relax}$, the characteristic time scales for a bound cluster, therefore, may further include the crossing times $t_{cr,i}=\{ R/v_t, R_h/v_t, r_t/v_t\}$, where $v_t$ represents the mean thermal velocity of the stars in the cluster. These considerations generalize (\ref{EQN_fN0}) to
\begin{eqnarray}
\frac{dN}{dt}=-f_N N F(t_{relax},t_{cr,i})
\label{EQN_fNa}
\end{eqnarray}
with accompanying loss of total energy
\begin{eqnarray}
\frac{dH}{dt}=f_H H F(t_{relax},t_{cr,i}).
\label{EQN_fEa}
\end{eqnarray}

Early attempts to include energy loss use order of magnitude scaling arguments \citep{spi40} or by explicit calculation of energy fractions in the tail of the velocity distribution \citep{hoe58,kin58,oss08}. This approach seems justified in the KM morphology, but requires careful estimation of the numerically small fractions involved, that represent a singular perturbation away from thermal equilibrium. The early estimates of \cite{kin58} and \cite{hoe58} give lifetimes shorter than $135t_{relax}$, but still much larger than what is found in numerical simulations by a few hundred percent due to an underestimate $f_H<f_N$. Present numerical estimates of the time to core-collapse range from $t_{ev}=10-20t_{rh}$ in terms of the relaxation time associated with the half-mass radius, $r_h$, that are considerably smaller than the \cite{amb38} evaporation time{,} by Fokker Planck and N-body simulations \citep{spi72,coh80,ina85,lee87,spu96,dru99,heg03}. Here, relatively longer times are found in King models with medium and low central concentrations \citep{qui96} and intermediate times in a Plummer model \citep{che90}, while short times are obtained in multi-mass models with broad mass distributions \citep{che90,joh93} with minor dependency on velocity anisotropies \citep{tak95}. 

{In this paper, we shall formulate the evaporative lifetime in terms of the coefficients $(f_N,f_H)$, derive (\ref{EQN_t04}) and determine its sensitivity to energy loss, going beyond that predicted by (\ref{EQN_fev}) alone. To pursue this, we first identify a critical number of stars, for evaporation to be limited by relaxation.}

\section{Relaxation-limited evaporation when the number of stars is large}

{At a given location in the cluster,} the {Boltzmann distribution} $n(E)=N^\prime\beta e^{-\beta E}$ for the kinetic energy $E$ (equal to the total energy as measured by a local observer) is the limit for detailed balance in a {local} thermal state,  where $N^\prime=\int_0^\infty n(E)dE$ denotes the total number of particles {in a local neighborhood}, $\beta = 1/k_BT$ the inverse temperature and $k_B$ Boltzmann's constant. Relaxation of small perturbations away from the Boltzmann takes place on the relaxation time scale $t_{relax}$ associated with small angle scattering in our present approximation. The normalized distribution $\bar{n}=e^{\beta E}n$ evolves to a uniform distribution in $E$. The process of relaxation for $\bar{n}$ is therefore equivalent to a diffusion process in a medium with uniform affinity (a vanishing chemical potential throughout), whereby $\partial_t \bar{n} = t_{relax}^{-1} \partial_E^2 \bar{n}$, i.e., 
\begin{eqnarray}
\partial_t n(E,t) =  t_{relax}^{-1}\left(\partial_E^2n-\beta \partial_En\right).
\end{eqnarray}

Evaporation is a process in which particles of energy $E>E_c$, {where $E_c$ depends on the location in the cluster,} escape on the time scale of the crossing time, $t_{cr}$, set by the size of the system, that provides an additional relaxation time for radiation leaking out to infinity. With the scaling $x=E/E_c$, $\alpha=E_c/k_BT$, 
\begin{eqnarray}
k_BT=\frac{\int_0^{E_c} En(E)dE }{ \int_0^{E_c}n(E)dE},
\label{EQN_kT}
\end{eqnarray}
we are led to consider the diffusion equation
\begin{eqnarray}
\partial_t n =t_{relax}^{-1} \left( \partial_x^2 n - \alpha \partial_x n\right) - t_{cr}^{-1} U(x-1) n,
\label{EQN_e2}
\end{eqnarray}
where $U$ denotes tthe Heaviside function. About a thermal distribution, Generally, the evolution of $n=n(x,t)$ depends on both timescales 
$(t_{relax},t_{cr})$. The case of $t_{relax}>> t_{cr}$ represents slow relaxation, relative to which evaporation on the time scale $t_{cr}$ is  
essentially immediate. It will be appreciated that (\ref{EQN_e2}) is similar in form to an inhomogeneous Kompaneets equation. The bound for this asymptotic result to hold allows by balancing the second and third terms in (\ref{EQN_e2}), i.e., 
\begin{eqnarray}
\frac{t_{relax}}{t_e}\simeq \alpha^2
\label{EQN_efs}
\end{eqnarray}
with a corresponding critical value $N=N_c$ satisfying $N_c\simeq 218 \log N_c$, i.e., 
\begin{eqnarray}
N_c\simeq 1600
\label{EQN_Nc}
\end{eqnarray}
by (\ref{EQN_tc}), (\ref{EQN_tr}) and the \cite{amb38} value $\alpha\simeq 6$. Thus, evaporation of clusters with $N>>N_c$ is limited by relaxation,  which justifies the starting point (\ref{EQN_fN0}-\ref{EQN_fev}) and hence (\ref{EQN_fNa}-\ref{EQN_fEa}) with
\begin{eqnarray}
F^{-1} = t_{relax}
\label{EQN_Frx}
\end{eqnarray}
and the corresponding Kelvin-Helmholtz time scale
\begin{eqnarray}
t_{KH}=f_H^{-1}t_{relax}.
\label{EQN_tKH}
\end{eqnarray}
The parameter range for relaxation-limited evaporation (\ref{EQN_Frx}) is amply satisfied by the globular clusters of the Milky Way with $N=10^{4-6}$. Young open star clusters in the disk of the Milky Way can be different with $N$ as low as a few hundred. For these systems, a full equation of the type (\ref{EQN_e2}) must be used.

A gradual departure away from the linear scaling of evaporative lifetimes $\propto t_{relax}$ has been observed in N-body simulations by \cite{bau01} in the range $N=10^2-10^4$, which we here identify with the low-N regime of (\ref{EQN_e2}) defined by (\ref{EQN_Nc}), where evaporative lifetimes depend on $t_{relax}$ and $t_{cross}$.  \cite{bau01} models the low-N scaling with
\begin{eqnarray}
F^{-1}=t_{relax}^{1-\beta} t_{cr}^{\beta} =  \left(\frac{6\ln N}{N}\right)^{\beta} t_{relax},
\label{EQN_FB}
\end{eqnarray}
where $\beta \simeq 0.25$. However, Eqn.(12) in \cite{bau01} from which (\ref{EQN_FB}) derives, does not recover the Boltzmann distribution in the absence of evaporation (defined by $t_{cr}\rightarrow \infty$). Hence, it fails to predict a corresponding critical value in $N$ that explains this transition. By (\ref{EQN_Nc}), we are able to ascertain that the discussions of \cite{bau01} (and \cite{lam05a,lam05b,gie08}) pertains strictly to low-$N$ asymptotic behavior in $N<N_c$ with no bearing on high-$N$ asymptotic behavior relevant to the present-day globular clusters with KM morphology in the Milky Way (with most having relatively small tidal parameters (\ref{EQN_t03})). A similar conclusion is given by \cite{mcl08}.

The existence of a transition to relaxation-limited evaporation in the large N limited has been anticipated but not specified in \cite{bau01,mcl08}. The above determines the transition to be around (\ref{EQN_Nc}) for $N$. The precise value of $N_c$ is not critical, as the transition is rather smooth, as may be seen from numerical results in Fig. 4 of \cite{bau01}. 

{In what follows, we shall continue in the parameter range $N>>N_c$, where the evaporation process is governed essentially by $t_{relax}$.}

\section{Evaporation coefficients $f_N$ and $f_H$ of a Maxwell distribution}

For an isolated cluster, an individual star is bound to escape whenever its {energy} is positive, i.e., $e_{k}+u>0$, where $e_k$ denotes the kinetic energy and $u$ the gravitational binding energy to all other stars in the cluster. For star $i$, we have $u=\Sigma_{j=1}^N U_{ij}$ (sum not including $j=i$). By (\ref{EQN_U}) and the Virial Theorem, the mean kinetic energy of the escaping stars hereby satisfies \citep{amb38}
\begin{eqnarray}
\bar{e}_{k} >  -\bar{u} = - 2\bar{U} = 4 \bar{E}_k.
\label{EQN_AMB}
\end{eqnarray}
Correspondingly, the escape velocity is about twice the mean velocity of the stars \citep{amb38,spi40}. For a thermal distribution, it represents a fraction
\begin{eqnarray}
f_N=\frac{\int_2^\infty  e^{-\frac{3}{2}s^2} (4\pi s^2)ds} {\int_0^\infty e^{-\frac{3}{2}s^2} (4\pi s^2)ds}=\frac{1}{134.44}
\label{EQN_fNb}
\end{eqnarray}
described by a Boltzmann distribution, {where $s^2=E_k/\bar{E}_k$, $\bar{E}_k=\frac{3}{2}k_BT$}, where $T$ denotes the temperature.
Alternatively, we may identify the cluster with the collection of bound stars whose velocities are restricted to the truncated Boltzmann distribution. In this event,
\begin{eqnarray}
f_N^*=\frac{\int_2^\infty e^{-\kappa s^2} (4\pi s^2)ds}{\int_0^2 e^{-\kappa s^2} (4\pi s^2) ds}=\frac{1}{112.43},
\label{EQN_fNb2}
\end{eqnarray}
where $\kappa=1.452165$ represents the modified relation $\bar{E}_k=\kappa k_BT$ obtained from (\ref{EQN_kT}), i.e.,
\begin{eqnarray}
\frac{\int_0^2 s^2 e^{-\kappa s^2} (4\pi s^2)ds}{\int_0^2 e^{-\kappa s^2} (4\pi s^2) ds} = 1. 
\end{eqnarray}
The difference in (\ref{EQN_fNb}-\ref{EQN_fNb2}) may be considered representative for the uncertainty in the present thermodynamic approach, and we will proceed with both to keep track of uncertainties in any of the derived quantities.

The relaxation time (\ref{EQN_tr}) is the instantaneous time scale for the relaxation of a stellar system, given its current state. It defines the rate at which the high velocity tail in the Boltzmann distribution of escaping stars is replenished by upscattering of low velocity stars, giving rise to the evaporation rate (\ref{EQN_fNa}) and the instantaneous evaporation time (\ref{EQN_fev}), or 21\% less based on (\ref{EQN_fNb2}). The estimate (\ref{EQN_fev}) represents an extrapolation based on the initial conditions of the system, which otherwise evolves nonlinearly over multiple relaxation times. Because stellar systems are self-gravitating, their temperatures increase as they loose stars by evaporation. Therefore, (\ref{EQN_fev}) gives an upper bound for complete evaporation, not the true time to complete evaporation in a fully nonlinear evolution. 

In the process of evaporation, the system evolves also in response to energy loss due to positive {energy} of the escaping stars. A complete system of equations, therefore, includes energy balance. Analogous to previous arguments, the total rate of energy loss (`luminosity') changes the total energy $H$ of the cluster at a rate (\ref{EQN_fEa}) with 
\begin{eqnarray}
f_H=\frac{\int_2^\infty s^2e^{-\frac{3}{2}s^2} (4\pi s^2) ds}{\int_0^\infty s^2e^{-\frac{3}{2}s^2} (4\pi s^2) ds}=\frac{1}{28.75}.
\label{EQN_fEb}
\end{eqnarray}
Here, we identify $H$ with the total energy of the stars below the threshold (all the stars that make up the cluster) and $f_H$ with the 
fraction of total energy carried off by the evaporating stars, that replenish the tail above the threshold in each relaxation time period. 

Note that $f_H$ is calculated by the relative fraction in kinetic energy in the tail of the velocity distribution, representing a flow in momentum space from local interactions (in coordinate space) between ``low" velocity stars (below the threshold, in the cluster), where kinetic energy and potential energy are virialized. Thus, replenishing the tail above the threshold with kinetic energy is tightly correlated to $H$ defined by the stars below the threshold and {\em vice versa}. Following (\ref{EQN_fNb2}), a more precise definition is, therefore,
\begin{eqnarray}
f_H^*=\frac{\int_2^\infty s^2e^{-\kappa s^2} (4\pi s^2) ds}{\int_0^2 s^2e^{-\kappa s^2} (4\pi s^2) ds}=\frac{1}{23.74},
\label{EQN_fEb2}
\end{eqnarray}
which differs from (\ref{EQN_fEb}) by 17\%.

The system (\ref{EQN_fNa}-\ref{EQN_fEa}) is closed by (\ref{EQN_H}). Integration of the two ordinary differential equations gives
\begin{eqnarray}
\frac{T}{T_0}=\left(\frac{N}{N_0}\right)^{-\gamma},~~\frac{R}{R_0} = \left(\frac{N}{N_0}\right)^{\alpha},
\label{EQN_sc}
\end{eqnarray}
where
\begin{eqnarray}
\gamma= \frac{f_N+f_H}{f_N}=5.71,~~\alpha = \gamma+1=6.71
\label{EQN_gamma}
\end{eqnarray}
based on $\gamma=5.712$ from (\ref{EQN_fNb}) and (\ref{EQN_fEb}). As a ratio, the estimate (\ref{EQN_gamma}) is stable in the first two digits in view of $\gamma=5.736$ as follows from (\ref{EQN_fNb2}) and (\ref{EQN_fEb2}). The correlations (\ref{EQN_sc}-\ref{EQN_gamma}) are {\em independent} of the choice of evolution time scale $F$.

\begin{figure}[htbp]
\centering\includegraphics[width=170mm,height=60mm]{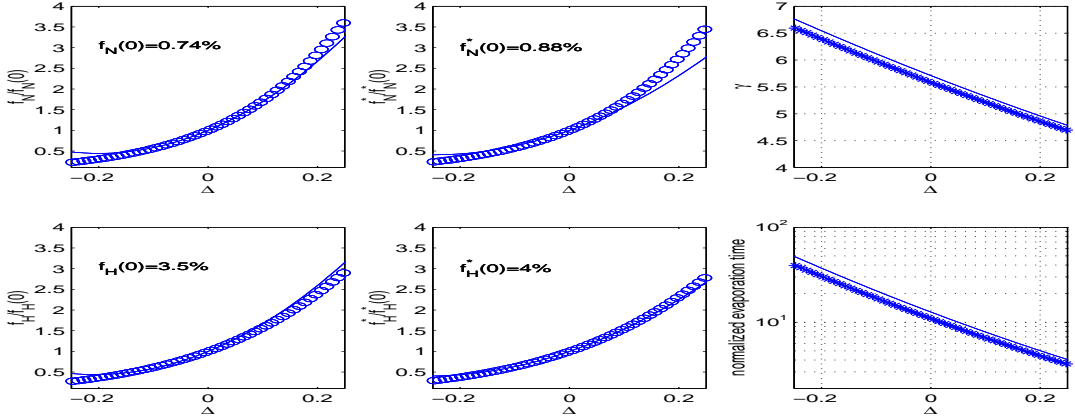}
\caption{({\em First and second column.}) Shown are the evaporation fractions $(f_N,f_H)$ and  $(f_N^*,f_H^*)$ as a function of the relative departure $\Delta$ from the nominal escape velocity ($circles$) over the range $\Delta=[-\frac{1}{4},\frac{1}{4}]$, including quadratic approximations thereto ($continuous$ $lines$), which differ by about 20\%. ({\em Third column.}) The resulting $\gamma=1+\frac{f_H}{f_N}$ 
(\ref{EQN_gamma}) are essentially the same for both pairs of fractions ($continuous$ and $thick$ lines, respectively). The normalized evaporation time $\frac{1}{qf_N}$ in (\ref{EQN_cc}) shows a steep dependency on $\Delta$ due to the faster than exponential decay in the tail of the Boltzmann distribution. }
 \label{fig_delta}
\end{figure}

As illustrated by (\ref{EQN_fNb}), (\ref{EQN_fNb2}), (\ref{EQN_fEb}) and (\ref{EQN_fEb2}), evaporation is a singular perturbation away from thermal equilibrium which introduces high sensitivity (of the coefficients $(f_N,f_H)$) to the critical escape velocity, even though some dimensionless results are relatively stable such as (\ref{EQN_gamma}). For example, a variation of +1\% in the threshold defining the tail gives rise to a variation of about +10\% in $t_{ev}$ (\ref{EQN_cc}). For small changes $\Delta$, we have 
\begin{eqnarray}
f_N(\Delta)=\frac{\int_{2-\Delta}^\infty  e^{-\frac{3}{2}s^2} (4\pi s^2)ds} {\int_0^\infty e^{-\frac{3}{2}s^2} (4\pi s^2)ds}=\frac{1+5.57\Delta+13.92\Delta^2+O(\Delta^3) }{135.44}
\label{EQN_fNb3}
\end{eqnarray}
\begin{eqnarray}
f_H(\Delta)=\frac{\int_{2-\Delta}^\infty s^2e^{-\frac{3}{2}s^2} (4\pi s^2) ds}{\int_0^\infty s^2e^{-\frac{3}{2}s^2} (4\pi s^2) ds}=\frac{1+4.73\Delta+9.45\Delta^2+O(\Delta^3) }{28.75}.
\label{EQN_fEb3}
\end{eqnarray}
Simular results hold for the fractions $(f_N^*,f_H^*)$, i.e., $f_N^*/f_N(0)\simeq 1+5.3781\Delta + 12.9308\Delta^2$, $f_H^*/f_H^*(0)\simeq 1+4.5419\Delta+8.6492\Delta^2$. Fig. \ref{fig_delta} shows both $(f_N,f_H)$ and $(f_N^*,f_H^*)$ as a function of $\Delta$. Corresponding to (\ref{EQN_fNb3}-\ref{EQN_fEb3}), the evaporation time satisfies
\begin{eqnarray}
\tau_{ev} (\Delta)=  12.82\left(1+\frac{0.05}{\ln N_0}\right)(1-5.01\Delta+14.13\Delta^2+O(\Delta^3))
\label{EQN_cc1}
\end{eqnarray}
and, similarly, $\tau_{ev}(\Delta) =  10.60\left(1+\frac{0.05}{\ln N_0}\right)(1-4.82\Delta+13.15\Delta^2+O(\Delta^3))$ in the approximation defined by $(f_N^*,f_H^*)$. Consequently, we have
\begin{eqnarray}
\frac{f_H}{f_N} = 4.71\left(1-0.8434\Delta + 0.23094\Delta^2+O(\Delta^3)\right)
\label{EQN_ee}
\end{eqnarray}
and similarly $\frac{f_H^*}{f_N^*} = 4.74\left(1-0.8460\Delta + 0.2150\Delta^2+O(\Delta^2)\right)$, which evidently is essentially the same for $(f_N,f_H)$ and $(f_N^*,f_H^*)$. We conclude that the impact of evaporation on energy is about five times larger than that on mass. 

\subsection{The evaporative lifetime of an isolated cluster}

For an isolated cluster, integration of (\ref{EQN_fNa}-\ref{EQN_fEa}) with (\ref{EQN_tr}), (\ref{EQN_sc}) and $F=t_{relax}^{-1}$ gives
\begin{eqnarray}
\mu^{q-1}\frac{d\mu}{d\tau}=-f_N\left(1+\frac{\ln\mu}{\ln N_0}\right)
\label{EQN_mu}
\end{eqnarray}
with $\tau=t/t_{relax}^0$, where $t_{relax}^0=\frac{N_0^\frac{1}{2}R_0^\frac{3}{2}}{6\ln N_0}$ denotes the initial relaxation time defined by $N_0$ and the initial radius $R_0$, $\mu=\frac{N}{N_0}$ the normalized particle number and $q=2+\frac{3}{2}\gamma=10.57$. A finite-time singularity solution obtains by ignoring the logarithmic term in (\ref{EQN_mu}),
\begin{eqnarray}
N=N_0 \left(1-\frac{\tau}{\tau}_{ev}\right)^\frac{1}{q},
\label{EQN_SN}
\end{eqnarray}
where $\tau_{ev}\simeq q^{-1}f_N^{-1}=12.82$. A next order approximation for $\tau_{ev}$ obtains by including a time average of the logarithmic term following (\ref{EQN_SN}). With $\ln\mu\simeq - \tau f_N$ for most of the time (away from the singularity), the time to complete evaporation extends to
\begin{eqnarray}
\tau_{ev}\simeq \frac{1}{qf_N}\left(1+\frac{1}{2q\ln N_0} \right)\simeq 12.82\left(1+\frac{0.05}{\ln N_0}\right).
\label{EQN_cc}
\end{eqnarray}
For all practical purposes, the evaporation time is about 13 times the instantaneous relaxation time, apart from a weak positive dependence on the initial particle number $N_0$ by no more than 5\%. 

Solutions of the type (\ref{EQN_SN}) are canonical for a broad range of radiation processes with generally different coefficients, that would include evaporation of relatively low N star clusters according to \cite{lam05b} on the basis of (\ref{EQN_FB}).  

The evaporative lifetime (\ref{EQN_cc}) serves as a lower bound to the core-collapse time of isolated clusters. Quite typically, scattering of stars occasionally has an impact parameter much smaller than $b_{min}=Gm/v^2$, in which case the interaction can be inelastic by tidal interactions that may lead to tidal capture and binaries formation (e.g. \citep{poo03}). This dissipative effect is ignored in idealized point-particle N-body simulations. While numerical simulations suggest that binary fractions of a 0.1\% can already produce an order of magnitude increase in the lifetime of the cluster \citep{jos01}, a detailed study identifying the threshold for a population of compact binaries to affect the lifetime of a cluster appears to be lacking.

\section{Stimulated evaporation by tidal fields}

In clusters bound to a host galaxy, the escape velocity will be perturbed by the tidal gravitational field of the galaxy, which generally facilitates stars to spill over across the Lagrange points $L_1$ and $L_2$, i.e., $e_{k}+u>-\delta$ following (\ref{EQN_AMB}), where $\delta>0$. Ab initio, $\delta$ depends on the direction relative to the line passing through $L_1$ and $L_2$ with an additional deflection due to the angular momentum of the globular cluster, as can be seen in direct N-body simulations (Fig. 7.4 in \cite{mey97}). The distance to $L_1$ and $L_2$ is given by the tidal radius \citep{spi87,che90,tak97}
\begin{eqnarray}
 r_t=a\left(\frac{M_c}{3M_G}\right)^\frac{1}{3},
 \label{EQN_rt}
 \end{eqnarray}
 where $M_G$ denotes the mass of the host galaxy and $a$ the separation of the cluster-galaxy system. 

Existing approaches to incorporate tidal effects commonly use the minimum energy for escape at $L_1$ \citep{spi87,lee87,che90,tak97,ros97}, effectively applying a spherically symmetric extension thereof to the tidal sphere. However, the probability for escape depends on the orientation relative to the anisotropic tidal field, and hence requires
averaging over the directions of last scattering events, here illustrated by Figs. \ref{fig_s1}-\ref{fig_s2} below. The possibility of long lived delays in escape \citep{bau03} conceivably modifies the escape probability, but this effect will be small in the limit $N>>N_c$.
 
As the relaxation time is smallest in the high density central region of a cluster, most of the escaping stars derive from high velocity jumps produced in the center \citep{joh93}. The tidal perturbation of the evaporation process can hereby be effectively modeled by studying the escape of stars that are emitted from a region about the center that is small relative to the tidal radius of the cluster (relative to the distance between the Lagrange points $L_1$ and $L_2$) in essentially randomly oriented directions; see also \cite{bau02,bau03}.  
 
The trajectory of a single star in the potential outside of the core of a globular cluster of mass $M_c=NM$ is the solution of a restricted three-body solution described by an effective potential in a Cartesian frame of reference $(x,y,z)=(a_c+r\cos\theta,r\sin\theta\cos\phi,r\sin\theta\sin\phi)$ with angular velocity $\Omega^2=\frac{M_G+M_c}{a^3}$ about the $z-$axis, given by
\begin{eqnarray}
\Phi=-\frac{M_c}{r_{c}}-\delta,~~\delta = \frac{M_g}{r_g} + \frac{1}{2}\Omega^2 \sigma^2-\frac{M_g}{a}\left[1 + \frac{1}{2}\left(\frac{a_c}{a}\right)^2\right],
\label{EQN_PHI}
\end{eqnarray}
where $\sigma=\sqrt{x^2+y^2}$ is the distance to the axis of rotation, and $r_{c}=\sqrt{(x-a_c)^2+y^2+z^2)}$ and $r_g=\sqrt{(x+a_g)^2+y^2+z^2}$ denote the distances of the star to the center of mass of the cluster and, respectively, the host galaxy. The latter have distances $a_{c}=a\frac{M_g}{M_g+M_c}$ and $a_g=-a\frac{M_c}{M_g+M_c}$ to the center of mass of the cluster-galaxy system, here at the origin of ($x,y,z$). We use units in which Newton's constant is equal to 1. Figs. \ref{fig_s1}-\ref{fig_s2} illustrate the tidal perturbations about the critical escape velocity, showing the critical onset of evaporation from the tidal sphere as a function of stellar velocity and anisotropy produced by the tidal field. There is further a gradual trend towards reduced anisotropy with increasing velocity. 

\begin{figure}[htbp]
\centering\includegraphics[width=140mm,height=100mm]{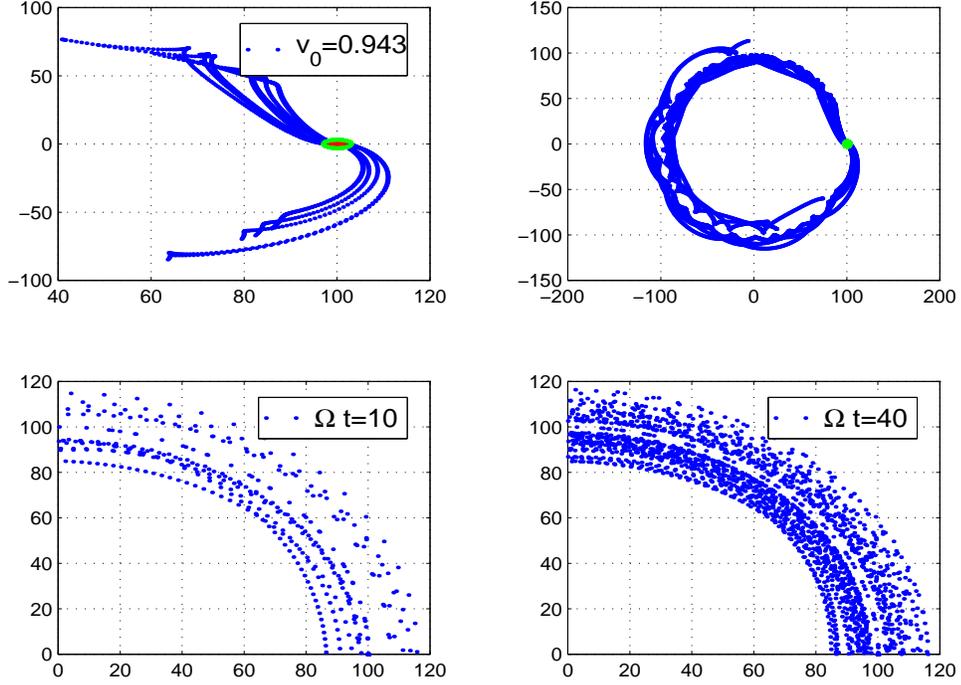}
\caption{($Top.$) Formation of $S-$shaped streams shown at two times $\Omega t = 10$ and $\Omega t=40$ in the corotating frame of the cluster by the trajectories $(blue)$ of 20 stars ejected uniformly in all directions from the cluster with mass $M=10^{-5}M_G$ with orbital angular velocity $\Omega$ around the host galaxy of mass $M_G$. The trajectories emanate from the center at $r_0=0.1r_t$, where $r_t$ is the tidal radius $r_t=1.5$ ($red$; $2r_t$ in $green$), relative to a distance of 100 from the center of mass of the cluster-galaxy system located at the origin. The results are shown for an initial velocity $v_0$ which is 0.943 times the escape velocity $v_0$ of the same cluster when viewed in isolation with no tidal field. ($Bottom.$) Transformed to the rest frame of the galactic center, the streams form extended tails as they would aggregate into diffuse rings in the absence of destruction by crossings with the galactic disk on the orbital time scale of the cluster ($not$ $shown$).}
   \label{fig_s0}
\end{figure}
\begin{figure}[htbp]
\centering\includegraphics[width=170mm,height=110mm]{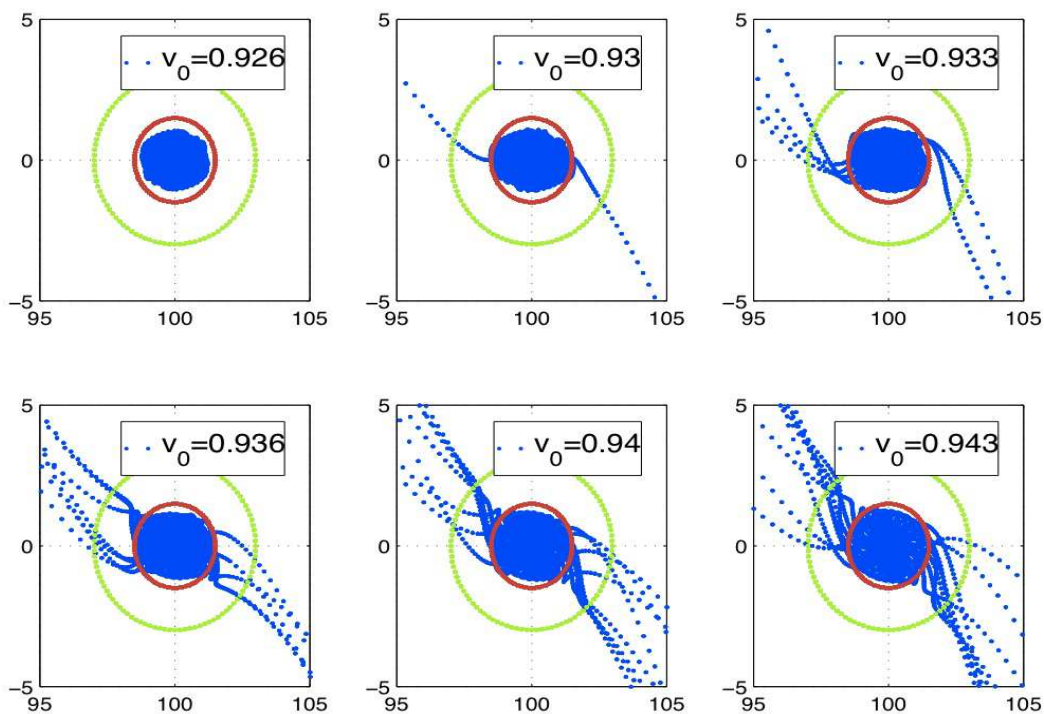}
\caption{Shown are the trajectories $(blue)$ of 30 stars ejected uniformly in all directions of a globular cluster similar to Fig. \ref{fig_s0} with initial velocities $v_0$ increasing by $0.5\%$ in each window, normalized to the critical velocity $\sqrt{\frac{2 M}{r_0}}$ of the same cluster in isolation with no tidal field. The tidal sphere forms an effective threshold for stars to escape. The results serve to illustrate the critical onset of anisotropic evaporation of high velocity stars in the tail of an extended velocity distribution in the presence of a tidal field.}
   \label{fig_s1}
\end{figure}
\begin{figure}[htbp]
\centering\includegraphics[width=170mm,height=110mm]{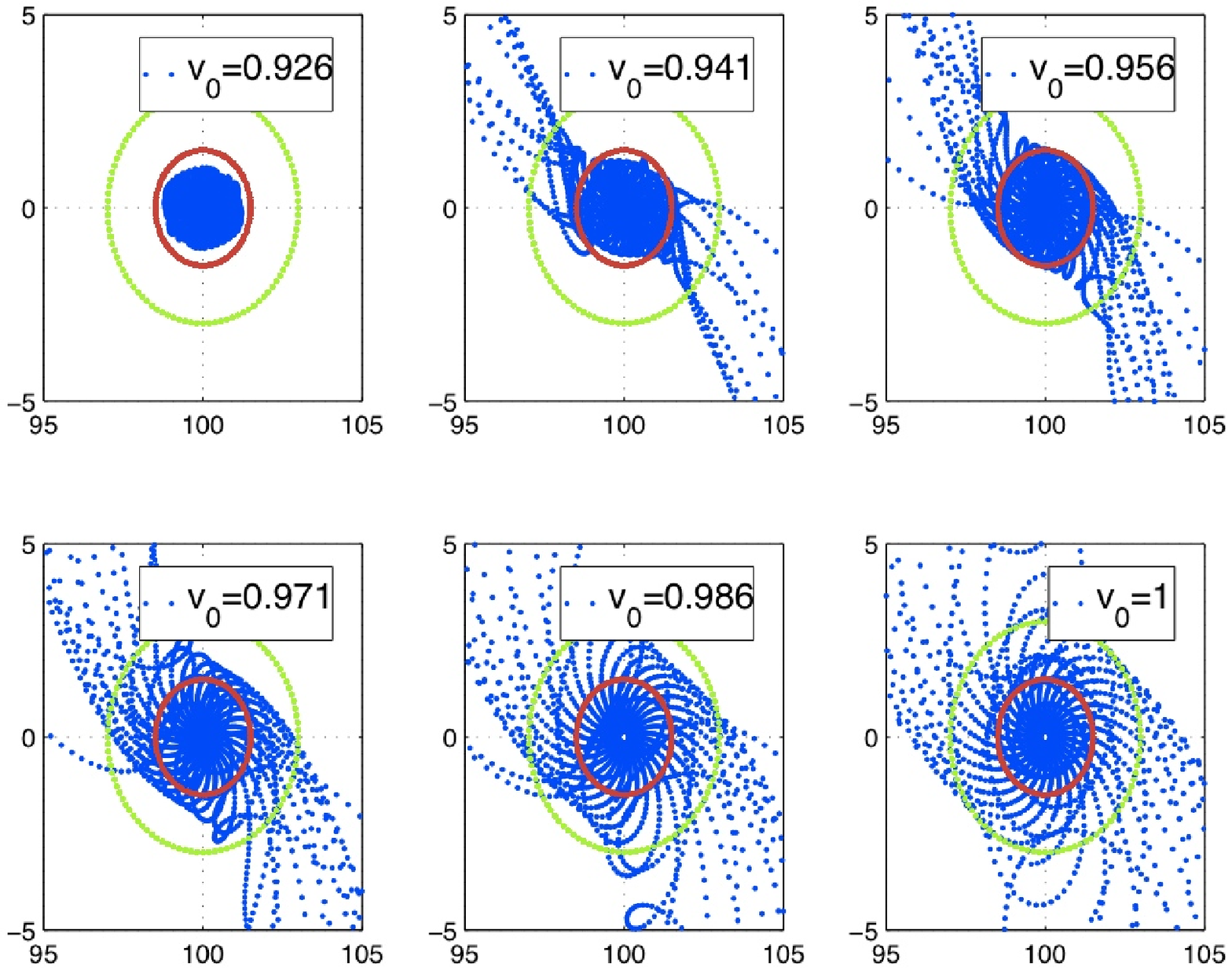}
\caption{Shown are the trajectories as Fig. \ref{fig_s1}, now with a choice of initial velocities $v_0$ increasing by $2.2\%$ in each window. The results serve to illustrate escape across the tidal sphere with an increasingly isotropic outflow of stars with increasing energies.}
   \label{fig_s2}
\end{figure}

\begin{figure}[htbp]
\centering{\includegraphics[width=80mm,height=60mm]{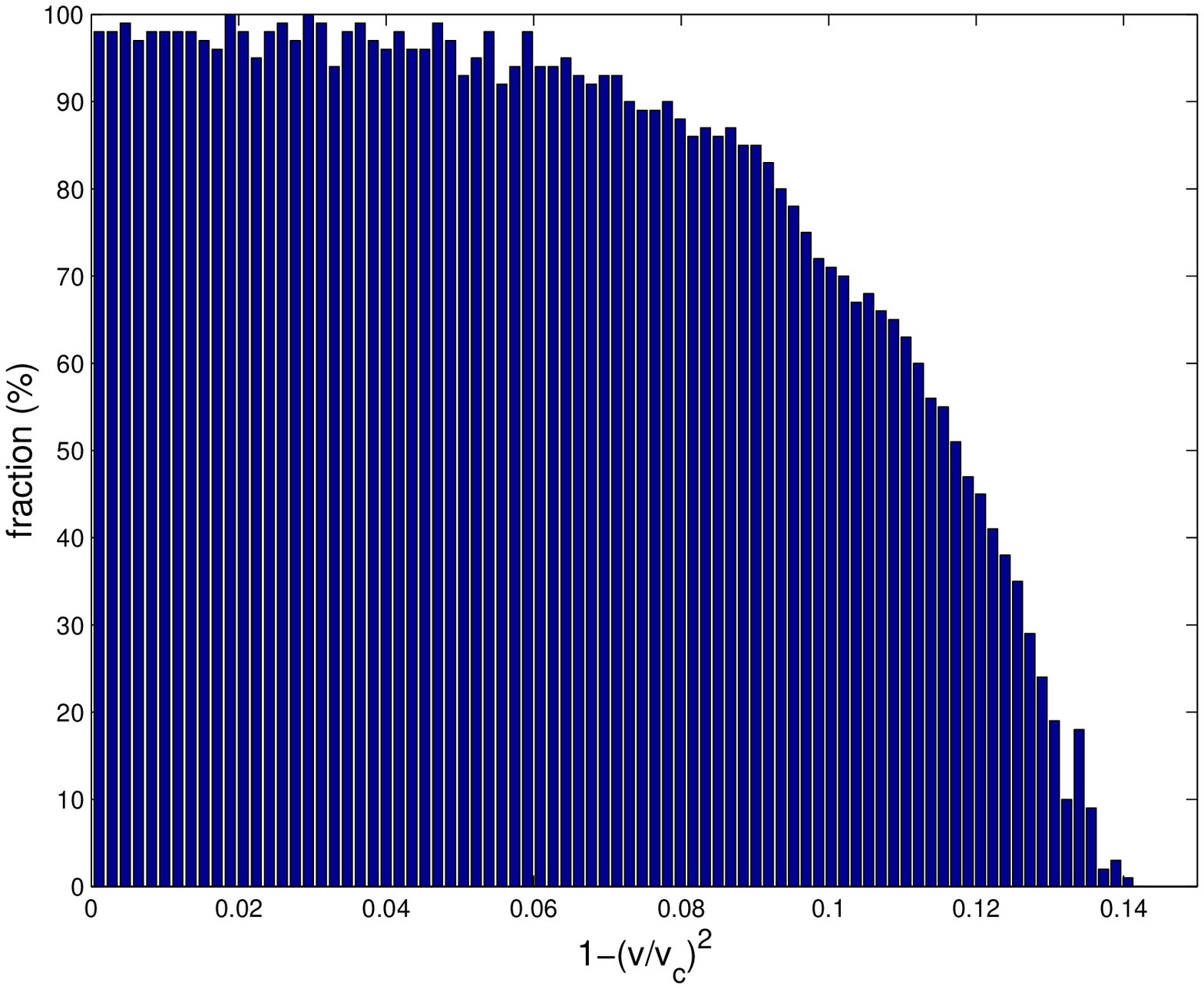}}
\centering{\includegraphics[width=80mm,height=60mm]{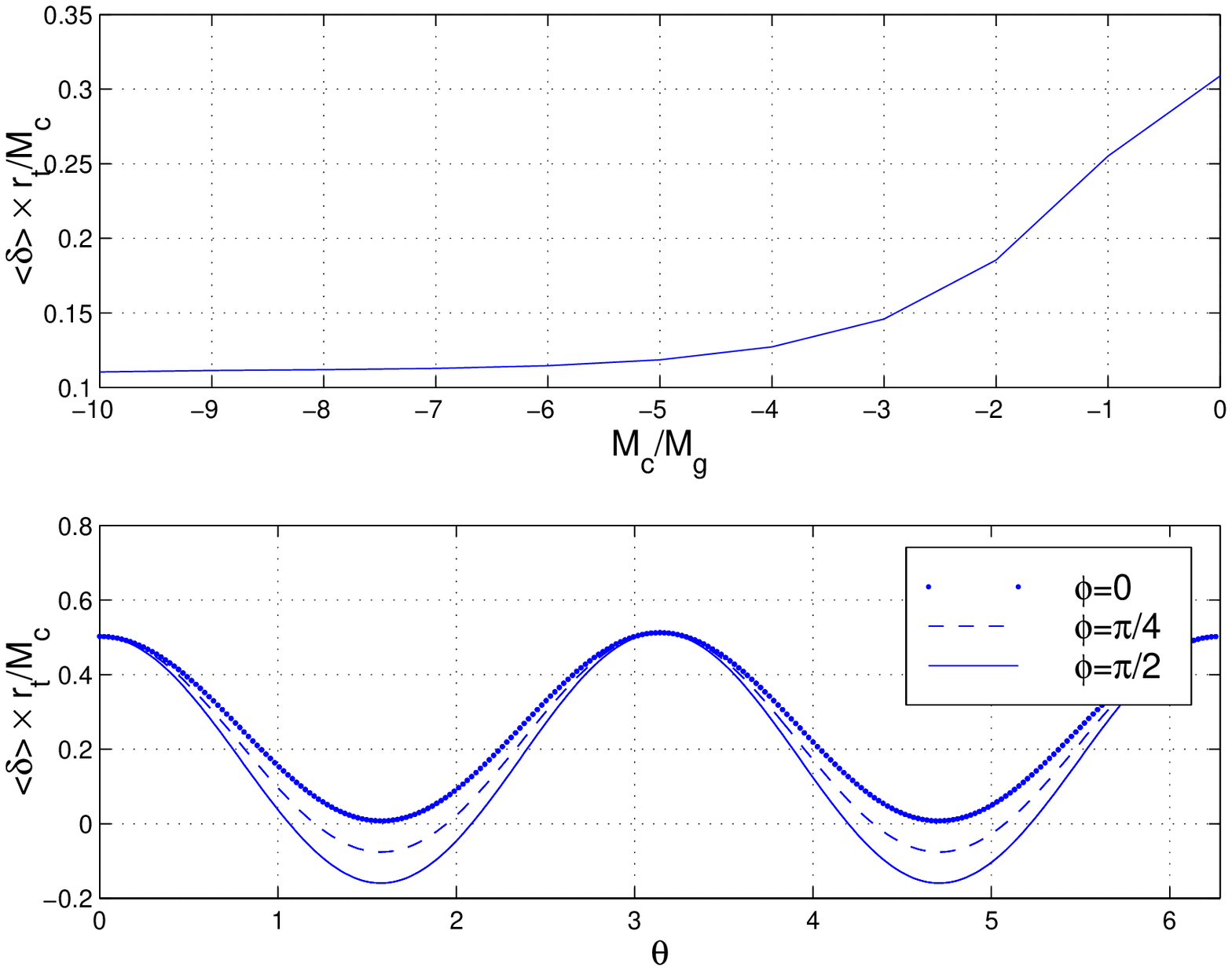}}
\caption{ ($Left.$) Shown is the fraction of escaping stars from a cluster with mass $M_c/M_g=10^{-5}$ $(\bar{\delta}\times r_t/M = 11.86\%)$ and tidal radius $r_t/a=1.5$, where $a$ denotes the to the host galaxy, computed according to Figs. \ref{fig_s1}-\ref{fig_s2}. The initial velocities $v$ are normalized to the critical velocity for escape $v_c$ if the cluster were isolated, whereby $1-(v/v_c)^2$ expresses the reduction in the potential energy barrier for escape due to the tidal field. The results show a reduced threshold in the kinetic energy for escape by about 14\% and a reduction of less than 12\% for a 50\% probability of escape. This sensitivity of a 50\% change in the flux of stars escaping in response to a 2\% change in $\Delta$ is consistent with the theoretical results summarized in Fig. \ref{fig_delta}. ($Right.$) Shown is the orientation averaged $<\delta>=\bar{\delta}$ kinetic energy for escape from a cluster bound to a host galaxy, relative to the kinetic energy for escape if the cluster were isolated. For typical cluster-to-galaxy mass ratios $M_c/M_g<<1$, the resulting grey-body factor $\Gamma=\bar{\delta}\times r_t/M_c$ asymptotes to about 11\%.}
   \label{fig_s3}
\end{figure}

The tidal interaction generally lowers the barrier for stars to escape from the tidal sphere by an energy perturbation $\delta$ that depends on the direction, as $\delta(r,\theta,\phi)$ has period $\pi$ in $\theta$ ($2\pi$ in $\phi$) with a pronounced amplitude in $\theta$ and minor variations in $\phi$ shown in Fig. \ref{fig_s3}. The Virial Theorem for the cluster (\ref{EQN_AMB}) now becomes 
\begin{eqnarray}
2\bar{E}_k+\bar{\Phi}=0,
\end{eqnarray} 
where $\bar{\Phi} = \bar{U}-\bar{\delta}$ with $\bar{U}$ and $\delta$ defined in (\ref{EQN_U}), respectively, (\ref{EQN_PHI}). It follows that $\bar{U}=-2\bar{E}_k + \delta$. The condition on the kinetic energy $e_k$ of a star to escape becomes $e_k  + u = e_k + 2\bar{U} - \delta >0$, giving a critical kinetic energy $4\bar{E}_k\left(1- \frac{\delta}{4\bar{E}_k}\right)$ reduced by the ratio of $\delta$ relative to the kinetic energy $4\bar{E}_k$ required for escape if the cluster were isolated. As the core of a globular cluster is much smaller than $r_t$, the escape of high velocity stars from it is effectively isotropic and the net escape rate obtains from averaging over all directions. The average of $\delta$ over the tidal sphere satisfies 
\begin{eqnarray}
\bar{\delta} = \frac{1}{4\pi} \int_0^\pi\int_0^{2\pi} \delta(r_t,\theta,\phi) \sin\theta d\theta d\phi = \Gamma \frac{M_c}{r_t},
\label{EQN_d1}
\end{eqnarray}
where the grey-body factor of the evaporation process satisfies
\begin{eqnarray}
\Gamma\sim 0.11-0.12
\label{EQN_grey}
\end{eqnarray}
in the limit of small ratios $M_c/M_g$ as shown in Fig. \ref{fig_s3}. With $\Delta = 2- 2\sqrt{1-\frac{\delta}{4\bar{E}_k}}$ and (\ref{EQN_U}), we arrive at 
\begin{eqnarray}
\Delta \simeq \frac{\bar{\delta}}{4\bar{E}_k} \simeq \Gamma \frac{R}{r_t}.
\label{EQN_Delta}
\end{eqnarray}
We are now in a position to calculate the modified evaporation rates due to a finite tidal radius $r_t$.

\subsection{The evaporative lifetime of a cluster in the tidal field of its host galaxy}

For the globular clusters of the Milky Way at the present epoch, the tidal ratio \ref{EQN_t03} assume rather moderate values broadly distributed around 0.1 \citep{bau09}. 
For this reason, we purse a regular perturbation of the evaporative lifetime in this tidal ratio, as opposed to a fully nonlinear treatment that would include tidal ratios of order one \citep{gie08}.

Following the previous section, we integrate the ordinary differential equations for mass and energy (\ref{EQN_fNa}-\ref{EQN_fEa}) with closure $(f_N,f_H)$ given by (\ref{EQN_fNb3}-\ref{EQN_fEb3}) and $\Delta$ in (\ref{EQN_t03}) expressed by (\ref{EQN_Delta}). For (\ref{EQN_ee}), we have, by the leading order solution (\ref{EQN_sc}),
\begin{eqnarray}
\frac{f_H}{f_N} = 4.71\left(1-0.097y + 0.0031y^2+O(y^3)\right)=a\left(1-by_0\mu^\frac{5+3a}{3} +O(y^2)\right),
\label{EQN_ee2}
\end{eqnarray}
where we take into account only the linear dependence on $y$, $a=4.71$, $b=0.097$, and $y_0$ refers to the initial value. Then the leading order solution for the energy evolution is
\begin{eqnarray}
H=H_0 \mu^{-a} e^{\frac{3ab}{5+3a}\mu^{\frac{5+3a}{3}}}\simeq H_0\left(\frac{N}{N_0}\right)^{-4.71} \left[1- c\left(\frac{N}{N_0}\right)^{\alpha-\frac{1}{3}}\right],
\end{eqnarray}
where $c=0.072y_0$, and similar to (\ref{EQN_sc}) we find
\begin{eqnarray}
\frac{T}{T_0} = \left(\frac{N}{N_0}\right)^{-\gamma} \left[1- c\left(\frac{N}{N_0}\right)^{\alpha-\frac{1}{3}}\right],~~
\frac{R}{R_0} = \left(\frac{N}{N_0}\right)^{\alpha} \left[1+ c\left(\frac{N}{N_0}\right)^{\alpha-\frac{1}{3}}\right]
\label{EQN_sct}
\end{eqnarray}
Similar to (\ref{EQN_sc}-\ref{EQN_gamma}), the correlations (\ref{EQN_sct}) are {\em independent} of the choice of evolution time scale $F$.

As when deriving (\ref{EQN_mu}) in the previous section with $\mu=N/N_0$, the evaporation time relative to the initial relaxation time follows by integration from (\ref{EQN_fNa}) with (\ref{EQN_tr}) and, now, (\ref{EQN_sct}) instead of (\ref{EQN_sc}),
\begin{eqnarray}
 \mu^{q-1}\frac{d\mu}{d\tau}\simeq -f_N^0\left[1+\frac{\ln\mu}{\ln N_0}-\frac{3c}{2}\mu^{\alpha-\frac{1}{3}}\right](1+0.64y_0\mu^\frac{2}{3} + 1.6y_0^2\mu^\frac{4}{3}),
\label{EQN_mu2}
\end{eqnarray}
where $f_N^0=\frac{1}{135}$.  

\begin{figure}[htbp]
\centering\includegraphics[width=100mm,height=70mm]{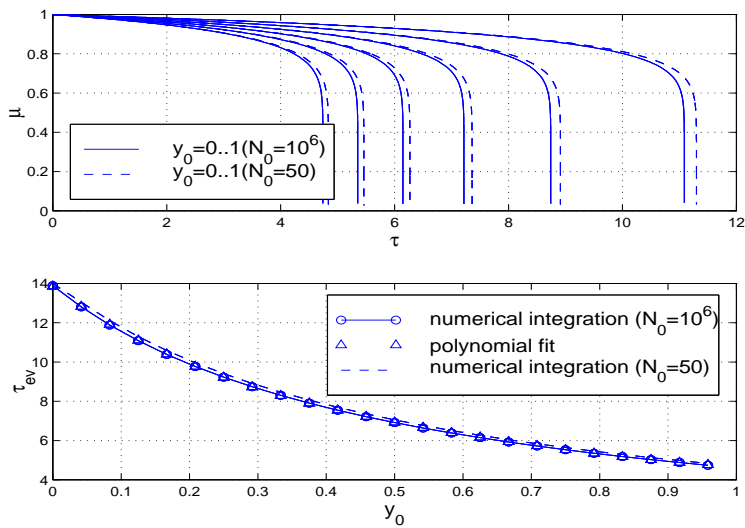}
\caption{($Top.$) Evolution of the number of stars $\mu=N/N_0$ for various initial tidal parameters $y_0$ according to (\ref{EQN_mu2}), where $\tau$ denotes the time in units of $t_{relax}$ defined by the virial radius $R$. $(Bottom)$ The slope to $\tau_{ev}$ is $-0.94$ at $y_0=0$, showing strong tidal sensitivity, here by numerical integration of (\ref{EQN_mu2}). The results are essentially the same for different $N_0$, showing that the process of relaxation-limited evaporation is essentially independent of $N$.}
\label{fig_delta2}
\end{figure}

Fig. \ref{fig_delta2} shown the result of numerical integration of (\ref{EQN_mu2}) using a canonical value $N_0=10^6$, where we infer a weak logarithmic dependence on $N_0$ by comparing results with $N_0=50$. With a polynomial fit to the tidal dependence, we arrive at (\ref{EQN_t04}), where we dropped the subscript 0 to initial values of $y$ and $N$, and where we dropped a factor $\left(1+0.002\ln N_6\right)^{-1}$ with extremely weak dependence on $N=N_6\times 10^6$. We emphasize that (\ref{fig_delta2}) refers to finite-time singularity solutions of single-mass clusters, as a regular tidal perturbation of isolated clusters in their pre-collapse state (e.g. Fig. 1 in \cite{bau02}). The detailed shape of the graph of $\mu(t)=N(t)/N_0$ in case multimass star clusters can be quite different with a continuously evolving mass function due to relatively fast evaporation of low-mass stars (e.g. Fig. 1 in \cite{bau03}). 

The numerical result shows a tidal sensitivity 
\begin{eqnarray}
\left.\frac{d\ln \tau_{ev}}{dy_0}\right|_{y_0=0}=-1.9,~~\left.\frac{d\ln \tau_{ev}}{dy_0}\right|_{y_0=1}=-0.70,
\label{EQN_tevt1}
\end{eqnarray}
giving rise to reductions up to a factor of a few when $y_0$ approaches 1 as shown in Fig. \ref{fig_delta2}. Very similar results follow from $(f_N^*,f_H^*)$. For young clusters with possibly $y_0>4$ such that $\Delta>0.2$, (\ref{EQN_mu2}) acquires contributions from $\Delta$ to higher order than those given in (\ref{EQN_fNb3}-\ref{EQN_fEb3}), which would require a further refinement of numerical integration. 

\subsection{Comparison with numerical simulations}

We may compare the above with existing approximations for the evaporation rate of stars in tidally truncated clusters \citep{spi87,tak97}, that are focused on $f_N$ with point wise extensions of the potential barrier at $L_1$, as opposed to spherical averaging (\ref{EQN_d1}) over the tidal sphere. This generally over-estimates the sensitivity to tidal interactions by a factor of about 4 given the amplitude of about $0.5$ in $\frac{r_t}{M}\delta$ as shown in \ref{fig_s3}. Furthermore, (\ref{EQN_tevt1}) may be compared with that resulting from the fit 
\begin{eqnarray}
t_{ev}=6.67\left( \frac{r_h}{r_t}\right)^{-1}
\label{EQN_fit}
\end{eqnarray}
 to numerical simulations in terms of the half-mass radius $r_h$ \citep{spi73,aug88}, satisfying
\begin{eqnarray}
\left.\frac{d\ln t_{ev}}{dy_0}\right|_{y_0=1}=-1.
\label{EQN_tevt2}
\end{eqnarray}

Our result given by (\ref{EQN_tevt1}) shows a generally $y-$dependent tidal sensitivity of the evaporation lifetime, that is overall consistent with the fit (\ref{EQN_tevt2}). The resulting evaporation lifetime (\ref{EQN_t04}) connects $y_0=1$ smoothly to the finite lifetime for isolated clusters ($y_0=0$). It will be noted that the fit (\ref{EQN_fit}) is restricted to a neighborhood of $y_0=1$, as it predicts an infinite lifetime for isolated clusters or, equivalently, clusters with very small tidal parameter (\ref{EQN_t03}). 

\cite{gie08} discuss a polynomial approximation to the instantaneous evaporation rate as a function of tidal radius. Beyond their cut-off of $y=0.05$, their Eqn.(7) predicts a logarithmic derivative of -1.5 for the evaporative lifetime to variations in the tidal ratio, similar to (\ref{EQN_tevt1}) derived above. 

\section{The remaining evaporative lifetimes of globular clusters of the Milky Way}

\begin{figure}[htbp]
\centering\includegraphics[width=175mm,height=140mm]{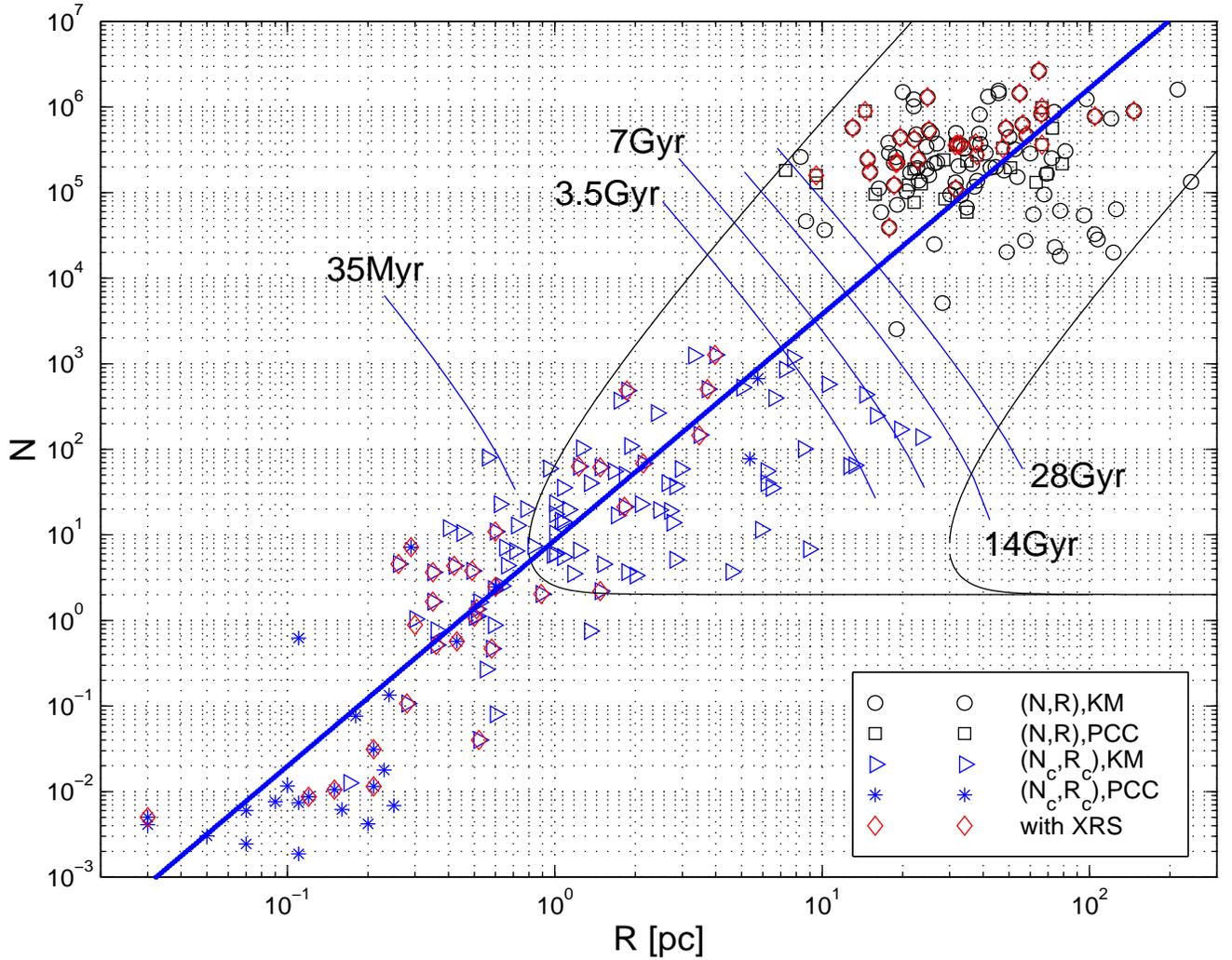}
\caption{Shown is the mass-radius distribution of 119 clusters in the Milky Way (from \cite{gne97}) in terms of $(N,R)$, where $N=M_h/M_\odot$ and $R$ denotes the radius at half mass $M_h$, differentiated in morphology for King Model (KM, $circles$) versus Post Core-Collapse (PCC, $squares$) (from \cite{che89}, adding PCC to NGC106 and  NGC4499). We add the distribution of inferred core parameters ($N_c,R_c$) in the Ansatz $N_c=(R_c/R)^3N$ (KM, $>$; PCC, $*$) which, for PCC, defines a lower bound on the associated $t_{ev}$. Curves of constant $t_{ev}$ are shown for 35 Myr, 3.5 Gyr, 7 Gyr and 14 Gyr. The 14 Gyr line separates the $(N,R)$ from $(N_c,R_c)$ without regard to morphology, consistent with the present age of the Universe and the age of the clusters. The average $t_{ev}$ for the clusters cores is about 1\% of that of the clusters. The combined distributions of $(N,R)$ and $(N_c,R_c)$ follows the {\em path of steepest descent} ($thick~blue~line$, slope 2.64) in $t_{relax}$ (two {\em curved black lines}), where the core-collapsed clusters continue this trend well below the $N=2$ model limit. A list of X-ray sources \citep{ver05} is included ($red~diamonds$), showing some preference towards high $N$ for a given $R$. Four clusters are between the curves of 14 and 28 Gyr, that will evaporate in the next Hubble time consistent with the evaporative destruction rate of \cite{agu88}.}
\label{fig_tcc}
\end{figure}

For globular clusters of the Milky Way, we recall that most have KM morphology, all satisfy (\ref{EQN_Nc}) and $y<<1$, by which (\ref{EQN_t04}) is applies with rather minor dependency on tidal effects. Independent arguments pointing to the same are given by \cite{mcl08,pri08}. 

Clearly, the $t_{ev}$ of the globular clusters exceed a Hubble time in agreement with their ages, where $t_{ev}=14$ Gyr separates the half-mass parameters on the right and the core-parameters ($N_c=(R_c/R)^3N,R_c)$ on the left. Overall, the distribution follows a trend along the paths of steepest descent of $t_{relax}$, where the core-collapsed clusters continue this trend well below the model limit $N=2$. 

Fig. \ref{fig_tcc} also shows that there is no correlation between the X-ray luminous binaries mentioned in the introduction and globular clusters with or without prior core-collapse, except that clusters with these X-ray sources tend to be of high mass for a given cluster radius. If (compact) primordial binaries contribute to a PCC morphology (e.g., \cite{gao91}), which is unclear at present, they are unlikely to contribute to the X-ray sources. Instead, the X-ray sources are likely produced in the course of stellar relaxation by tidal captures \citep{poo03,ben06} with no need for a primordial binary population.  

\section{Cluster-tail correlations produced by evaporation} 

Evaporation of stars modulated by tidal fields give rise to anisotropic outflows with bipolar morphology, commonly referred to as tidal tails. Relaxation limited evaporation provides a launching mechanism that predicts a correlation between the velocity of growth of these tidal tails and the velocity dispersion in the cluster as defined by its temperature \citep{kup08a,kup08b,kup09}. We here quantify this correlation suitable for detailed observational studies. 

Following (\ref{EQN_fNa}-\ref{EQN_fEa}), the mean kinetic energy $\bar{e}_k(r)$ of the escaping stars at distance $r$ are related to the mean kinetic energy of the stars in the cluster by
\begin{eqnarray}
\bar{e}_k(r) = \left(\frac{f_H}{f_N}-4\right)\bar{E}_k+\frac{mM}{R}y\frac{r_t}{r},
\label{EQN_me}
\end{eqnarray}
where we encounter the evaporation efficiency
\begin{eqnarray}
\frac{f_H}{f_N}-4 = 0.71\left(1-0.097y + 0.0031y^2+O(y^3)\right)
\label{EQN_ee3}
\end{eqnarray}
by (\ref{EQN_ee2}). 

Clusters with $y<1$ will produce a bipolar outflow of stars that is amenable to detailed observations as mentioned in the introduction. These tidal tails carry a record of the most recent evaporation process by the number and kinetic energies of their stars as measured relative to the center of mass of the cluster. The planned and upcoming LSST, BigBOSS, JASMINE and GAIA offer unique prospects for extracting kinematic information in a wide field of view around globular clusters with angular resolutions around 10 $\mu$ as, resolving velocities of individual stars to about 1 km/s at visual magnitudes typical for globular clusters. To test (\ref{EQN_me}), we propose (1) resolving the kinetic energy in tidal tails outside the tidal sphere, (2) the velocity dispersion in the cluster and (3) comparing for the ratio of the two with the predicted value 
\begin{eqnarray}
\bar{e}_k\ge 0.71 \left(1-0.097y +O(y^2)\right)\bar{E}_k,
\label{EQN_p}
\end{eqnarray}
were equality is approached at large distances away from the cluster, when the gravitational binding energy to the cluster can be neglected. Following \S2, $R=1.30r_h$ in $N\bar{E}_k=\frac{GM_c^2}{4R}$ gives a mean (three-dimensional) velocity dispersion in the cluster,
\begin{eqnarray}
\sigma=\sqrt{\frac{0.19 GM_c}{r_h}},
\label{EQN_V1}
\end{eqnarray}
and a scale for bipolar outflow 
\begin{eqnarray}
\bar{V}_t\ge \sqrt{\bar{V_t^2}}=0.84\left(1-0.049y+(y^2)\right)\sigma
\label{EQN_V2}
\end{eqnarray}
by (\ref{EQN_p}). The velocity of the tidal {\em front} may be larger than $\bar{V}_t$, as it is determined by the fastest escaping stars. For instance, \cite{leo00} report for NGC 6254 a projected expansion velocity 7 km s$^{-1}$ of tidal material at a distance of $r=$150 pc ($D=$4.1 kpc) and a projected velocity dispersion 6.6 km s$^{-1}$ in the cluster, were it is noted that the gravitational binding energy to the cluster is negligible at this distance. Evidently, an expansion velocity exceeding the velocity dispersion in the cluster satisfies (\ref{EQN_V2}). The correlation (\ref{EQN_p}) is relatively insensitive to various cluster parameters, such as the total mass, which otherwise is not well defined in view of an uncertain but possible substantial population of white dwarfs \citep{bau03}. 

In Table 1 we give an illustrate sample of our model predictions, where we assume a mean stella mass $\bar{m}=\frac{1}{3}M_\odot$ to derive $N=M_c/\bar{m}$ for each cluster. We focus on tidal tails created over the period $T=T_710^6$ yr of a few times ten million years since the last disk crossing, to circumvent the uncertain fate of extended tidal tails shown in Fig. \ref{fig_s0} following the most recent disk crossing. Scaled to one-quarter of the period $P=P_6$ Myr of the orbit around the galactic center, the associated outflow of evaporating stars satisfies
\begin{eqnarray}
n\simeq  \frac{PN f_N }{4t_{relax}} = 0.275\times N_5 P_6 \left(\frac{t_{rh}}{10^9\mbox{~yr}}\right)^{-1}
\label{EQN_n}
\end{eqnarray}
with $N=10^5N_5$.

\begin{table}[h]
{\bf TABLE I.} Data on selected globular clusters derived from the catalogue of \cite{har96} (revision 2010) and model predictions for the outflow velocity $V_t$ from $\sigma$ given by (\ref{EQN_V1}), the number of escaping stars $n$ per one-quarter orbital period by (\ref{EQN_n}) and the projected angular separation $\alpha_0$ between the leading and trailing tails inside and outside the orbit derived from (\ref{EQN_j}) using positional data. Where available, a comparison with observational data is included.\\ \\
\centerline{
\begin{tabular}{llrrrrrrrrrll}
\hline
Name$^a$ & $D^b$ & $M_c$ & $r_t$ & $y_h$ & $t_{rh}$ & $A$ & $r_h$ & $\sigma$ & $\bar{V}_t$ & $n $ & $\alpha^e_o$\\
 &   {\tiny kpc}     &  {\tiny $10^6 M_\odot$} & {\tiny pc}   &  & {\tiny Gyr}    &         & {\tiny pc}       &  {\tiny km/s}	   &   {\tiny km/s}  &  & {\tiny arcmin} \\
\hline
\hline
NGC 104$^1$ 		& 4.5(7.4)	 & 1.45       	&   64                           &   0.065 	& 3.5   & 3.7 & 4.1  & 16   & 14   & 140 	& 120($\sim$ 100)$^f$ \\
Pal 4$^1$         		&  109(111)  &  0.0541 	&   100                 	&  0.162   	& 2.6   & 2.0 & 16  &  1.6 & 1.3 	& 161  	&  1 \\
Pal 5$^1$         		&  23.2(18.6)&  0.0284	&   101                  	&  0.182   	& 6.6   & 2.1 & 18  &  1.1 & 0.9 	& 6      	& 0.4($<30$)$^{f,h}$\\
Pal 14$^1$       		& 73.0(71.6) &  0.0200	&   118                   	&  0.219   	& 10    & 1.5 & 26   & 0.8 & 0.6 & 23     	& 0.06\\
NGC 5139$^1$ &  5.2(6.4)&   2.64   		&   104                    	&  0.073    	& 12    & 1.2 & 7.6  & 16  & 14	& 64	  	& 75($\sim$ 100)$^f$\\
NGC 5904$^1$  &  7.5(6.2) & 0.834  		&   52                  &   0.074 		& 2.5   & 5.3 & 3.9  & 13   & 11 	& 87	   	&  43$<60$)$^h$ \\
NGC 4590$^1$ &  10.3(10.2) & 0.306  	&   89                      	&   0.051   	& 1.9   & 6.4  & 4.5 & 7.3   & 6.1 & 82    	&  22($<60$)$^h$\\
NGC 5466$^1$ &   16.0(16.3)& 0.133  	&   162                 	&   0.066   	& 5.7   & 2.5  & 11  & 3.1  & 2.6 	&  29		&   4($<60$)$^g$\\
NGC 6254$^1$& 4.4(4.6)	& 0.225   		&   30                    	&  0.084    	& 0.8   & 18 & 2.5   & 8.4   & 7.0 &  51	& 70($\sim$ 100)$^f$\\
NGC 6121$^1$ &1.7$^d$(5.9)& 0.225  	&   19                      	&  0.112    	& 0.9   & 23  & 2.1  & 9.0    & 7.6 & 93	& 324 &\\
NGC 6752$^2$&4.0(5.2) & 0.364  		&   52                  	&   0.042   	& 0.7   & 17  & 2.2  & 11    & 9.5 &  96	& 46  &\\
NGC 7078$^2$&10.4(10.4)& 0.984 	 	&   62                           &   0.049   	& 2.1   & 7.1 & 3.0  & 16    & 13 	&  301	 & 8.7 & \\
\hline
\end{tabular}
\label{TABLE_1}}
$^1$ King morphology \citep{che89}\\
$^2$ Post core-collapse morphology \citep{che89}\\
$^a$ NGC 104=47 Tuc, NGC 5139=$\omega$ Centauri, NGC 5904=M5, NGC 4590=M68, NGC 6121=M4, NGC 6254=M10, NGC 7078=M15\\
$^b$ Distances to the Sun (galactic center) with uncertainties of $6\%$ \citep{cha07}\\
$^c$ \cite{che89}\\
$^d$ \cite{ric04}\\
$^e$ Model predictions for projected angular separation between leading and trailing tidal streams (observed values, if known) from the position of the globular cluster assuming circular motion and $M_g=3\times 10^{11}M_\odot$.\\
$^f$ \cite{leo00}\\
$^g$ \cite{ode04,bel06,fel07}\\
$^h$ \cite{jor10}
\mbox{}\hskip0.01in
\end{table}

By angular momentum conservation, bipolar outflows emanating from the two Lagrange points $L_1$ and $L_2$ deflect into leading and trailing streams inside and outside the orbit of the cluster \citep{heg01}. We express it by a correlation of the radial separation to the temperature of the cluster according to an angular separation
\begin{eqnarray}
\alpha = \frac{\Delta r}{D_{gc}} = 3.4 \frac{\sigma}{D_{gc}\Omega} = 0.55~ \sigma_1 D_4^{1/2}~\mbox{deg},
\label{EQN_j}
\end{eqnarray}
where $\Omega=\sqrt{\frac{M_g}{D^3}}$ with $M_g=3\times 10^{11}M_\odot$ \citep{ode97} denotes the orbital period of the cluster about the galactic center at a distance $D_{gc}=D_4$ 10 kpc and $\sigma=\sigma_1$ 1 km $s^{-1}$. The projected angular separation follows from the known positions of the globular clusters, shown in the last column of Table 1 in the approximation of circular motions. 

\section{Conclusions and observational outlook}

Stellar clusters are open systems that radiate stars and energy by evaporation{, mostly by diffusion in momentum space subject to a singular perturbation by rare but finite jumps due to hard 2- or possibly 3-body interactions}. We identify a parameter range for relaxation-limited evaporation applicable to clusters in their pre-collapse phase (generally of KM morphology) with $N>>N_c$ and small to moderate tidal ratios $y<1$, and develop a model for the evaporative lifetime for single-mass clusters. {For a Maxwell velocity distribution that extends beyond the escape velocity, evaporation is} {\em bright} as characterized by an inequality between the Kelvin-Helmholtz (\ref{EQN_tKH}) and Ambartsumian-Spitzer time scales (\ref{EQN_fev}), satisfying
\begin{eqnarray}
t_{KH}<t_{AS}
\label{EQN_KHAS}
\end{eqnarray}
due to $f_H>f_N$, wherein evaporation is largely driven by energy-loss rather than mass loss. 
The evolution isolated clusters, as a reference to more realistic modeling, is hereby described by a finite-time singularity solution
\begin{eqnarray}
\{N, R, T \}\propto  \left(1-\frac{\tau}{\tau}_{ev}\right)^{\{0.0905,~0.6351,~-0.5405\}}.
\label{EQN_RT}
\end{eqnarray}
It shows that the temperature and the radius evolutions (\ref{EQN_RT}) are relatively less shallow than (\ref{EQN_SN}), but still evolve only moderately away from the time of complete evaporation according to (\ref{EQN_RT}). Here, the indices $0.0905$ and 0.6351 in (\ref{EQN_RT}) are close to the values 0.085 and 0.695, respectively, observed in N-body simulations \citep{heg03}. We do not attribute the relatively small $t_{ev}$ (relative to $t_{AS}$) to a radial stratification of the relaxation time with $f_H=0$ as proposed in \citep{joh93}. It also {at odds with} an older claim that $f_H<f_N$ (with commensurably $t_{KH}>t_{AS})$ by \cite{hoe58,kin58,kin58b}. Similarly, stars escaping by ejections from last scattering events from within the cluster should be contrasted with low velocity evaporation arising from diffusive escape through a low temperature outer boundary, commonly used in the Fokker-Planck approximations as if the cluster were opaque throughout up to its tidal boundary \citep{spi87}.

{The agreement of (\ref{EQN_t04}) with numerical simulations supports the idea that energy loss in accord with (\ref{EQN_KHAS}) is important. For a Maxwell velocity distribution, (\ref{EQN_t04}) is shorter than that based on particle loss alone by a factor of about three due energy loss (\ref{EQN_fEb}). It would be of interest to identify the various scattering processes that may give rise to a relatively large $f_H$, see e.g. \cite{pet70,ret79,joh93,mey01a,ash04}, and determine the implied shortening of the evaporative lifetime following \S5.1. These studies fall outside the scope of the present work, however. Instead, we focus on some observational tests in \S8.}

The presence of a tidal field modifies the net rate of evaporation by two effects. It generally enhances evaporation by lowering the threshold on the kinetic energy for stars to spill over the tidal radius, but the escape probably is attenuated by the generally anisotropic tidal field as defined by the Lagrange points $L_1$ and $L_2$ described by the grey body factor (\ref{EQN_grey}). For moderate strengths of the tidal field, the resulting evaporation rate can be expanded in a regular perturbation in the tidal ratio (\ref{EQN_t03}). 

In studying the globular clusters of the Milky Way, we note that they (1) amply satisfy the criterion for relaxation-limited evaporation (\ref{EQN_Nc}), (2) their tidal rations satisfy $y<<1$, (3) most of them are of KM morphology indicative of a pre-core collapse state. For this reason, the perturbative expansion (\ref{EQN_t04}) for their remaining relaxation-limited evaporative lifetimes is believed to be applicable, and it is found to be consistent with their current ages. 

The energy of the escaping stars is directly correlated to the temperature of the cluster (\ref{EQN_j}), and the outflows thus produced develop a bi-polar $S-$shaped morphology in the plane of the Lagrange points $L_1$ and $L_2$ and the orbit of the cluster. For those tidal tails where the $S-$shaped tail can be resolved observationally, the de-projected orbital displacement of the tails is representative for the outflow velocity by the correlations (\ref{EQN_V2}-\ref{EQN_j}).   

If periodically destroyed by disk crossings, these tails hereby re-appear on a time scale of 10 Myr with a length scale of a few times the tidal radius for clusters that are not too small in mass and velocity dispersions. Detailed observational studies are proposed to test (\ref{EQN_V2}) on a wide field of view around globular clusters using upcoming high resolution photometric and spectroscopic stellar survey instruments. The results may also be used for comparison with alternative mechanisms for inducing tidal tails, such by tidal heating or shocks in response to disk crossings \citep{heg01} and velocity anisotropy \citep{tak00}. We remark that destruction by tidal fields of an initial distribution might have been more severe around spiral galaxies than around the relatively larger elliptical galaxies, which tend to have relatively large numbers of globular clusters in proportion to their luminosity \citep{har81,bur10}. 

The upcoming large field of view photometric and spectroscopic instruments LSST, BigBOSS, GAIA and JASMINE seem to be ideally suited for detailed measurements on tidal tails in energy (\ref{EQN_V2}), stellar count (\ref{EQN_n}) and morphology (\ref{EQN_j}) to advance an observational test of the present thermodynamic model.

{\bf  Acknowledgment.} The author gratefully acknowledges detailed constructive comments from the anonymous referee, J. Scalo and H. van Beijeren.

\end{document}